\documentclass[fleqn,usenatbib]{mnras}

\usepackage{newtxtext,newtxmath}

\usepackage[T1]{fontenc}

\DeclareRobustCommand{\VAN}[3]{#2}
\let\VANthebibliography\thebibliography
\def\thebibliography{\DeclareRobustCommand{\VAN}[3]{##3}\VANthebibliography}

\usepackage{graphicx}	
\usepackage{amsmath}	
\usepackage{nccmath}

\usepackage{xcolor} 
\usepackage{soul} 

\usepackage[nolist]{acronym}    

\usepackage{booktabs}   

\usepackage{float}      
\usepackage{tablefootnote}
\usepackage[]{threeparttable}

\usepackage{subcaption} 
\usepackage{threeparttable}
\usepackage{multirow}
\usepackage{anyfontsize}

\newcommand{\kms}{\,km\,s$^{-1}$} 

\newcommand{\lsun}{\ifmmode{{\rm ~L}_\odot}\else{~L$_\odot$}\fi}
\newcommand{\Msun}{\ifmmode{{\rm ~M}_\odot}\else{~M$_\odot$}\fi}
\newcommand{\sqdeg}{\,deg$^2$}
\newcommand{\ujybm}{$\mu$Jy/bm\,}

\begin{acronym}
    \acro{AGN}      {Active Galactic Nuclei}
    \acro{ASKAP}    {Australian Square Kilometre Array Pathfinder}
    \acro{ATNF}     {Australia Telescope National Facility}
    \acro{EMU}      {Evolutionary Map of the Universe}
    \acro{LIRG}     {Luminous Infrared Galaxy}
    \acro{ORC}      {Odd Radio Circle}
    \acro{RM}       {Rotation Measure}
    \acro{SFG}      {star-forming galaxy}
\end{acronym}

\title[MeerKAT discovery of an ORC]{MeerKAT discovery of a MIGHTEE Odd Radio Circle}

\author[Ray P. Norris et al.]{Ray P. Norris$^{1,2}$,\thanks{E-mail: ray.norris@csiro.au}
B\"arbel S. Koribalski$^{1,2}$,
Catherine L. Hale$^{3}$,
Matt J. Jarvis$^{3,4}$,
Peter J. Macgregor$^{2,1}$,
\newauthor
A. Russell Taylor$^{5,4}$
\\
$^{1}$CSIRO Space and Astronomy, P.O. Box 76, Epping, NSW 1710, Australia \\
$^{2}$School of Science, Western Sydney University, Locked Bag 1797, Penrith, NSW 2751, Australia \\
$^{3}$Astrophysics, Department of Physics, University of Oxford, Keble Road, Oxford, OX1 3RH, U.K.\\
$^{4}$Department of Physics and Astronomy, University of the Western Cape, Bellville, Cape Town, 7535 South Africa\\
$^{5}$The Inter-University Institute for Data Intensive Astronomy (IDIA), Department of Astronomy, University of Cape Town, Private Bag X3, Rondebosch, 7701, South Africa 
}

\date{Accepted XXX. Received YYY; in original form ZZZ}

\pubyear{2024}

\begin{document}

\label{firstpage}
\pagerange{\pageref{firstpage}$-$\pageref{lastpage}}
\maketitle

\begin{abstract}
We present the discovery of a new Odd Radio Circle (ORC J0219--0505) in 1.2~GHz radio continuum data from the MIGHTEE survey taken with the MeerKAT 
The radio-bright host is a massive elliptical galaxy, 
which shows extended stellar structure, possibly tidal tails or shells, suggesting recent interactions or mergers. The radio ring has a diameter of 35\arcsec, corresponding to 114~kpc at the host galaxy redshift of $z_{\rm spec} = 0.196$. This MIGHTEE ORC is a factor 3--5 smaller than previous ORCs with central elliptical galaxies. 
The discovery of this MIGHTEE ORC in a deep but relatively small-area radio survey implies that more ORCs will be found in deeper surveys. 
While the small numbers currently available are insufficient to  estimate the flux density distribution, this is consistent with the simplest hypothesis that ORCs have a flux density distribution similar to that of the general population of extragalactic radio sources.
\end{abstract}






\section{Introduction}


\acp{ORC} are extragalactic circles of steep-spectrum radio emission, without any corresponding emission at other wavelengths, other than that of the host galaxy. They were first discovered by \citet{norris21a} in the Evolutionary Map of the Universe Pilot Survey \citep[EMU-PS1,][]{norris21b} using the \acl{ASKAP} \citep[ASKAP,][]{ hotan21}, and the first  detailed image of an ORC was made by \citet{norris22} using the MeerKAT telescope \citep{jonas16}.

\begin{figure}
    \includegraphics[width=8.5cm]{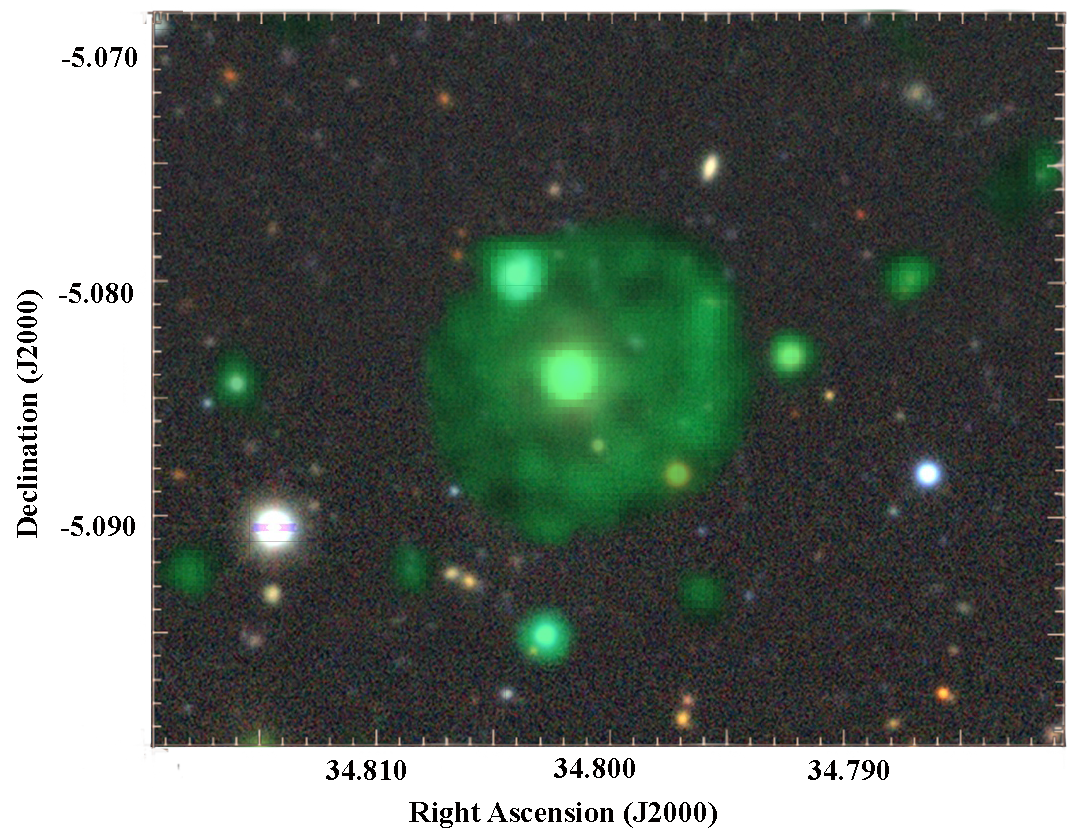}\hfill
    \caption{Composite image of ORC J0219-0505 consisting of the MeerKAT Stokes I image (green) superimposed on an RGB optical image from the  Legacy Survey  Interactive Sky Viewer. The transfer function for the  radio data of the field of view was adjusted for contrast, and assigned to green. The colour scheme for the optical data uses the default colour scheme for Legacy Survey DR9 data assigned by the Legacy Survey  Interactive Sky Viewer.  The two confirmed companion galaxies, just north and south of the ORC host, have bluish colours, but appear pale green here because of their radio emission. This figure is optimised to convey the structure of the ORC, and quantitative information should be taken from
    Figure \ref{fig:3bands} or from the MIGHTEE DR1 release data.}
    \label{fig:neworc}
\end{figure}

In the initial discovery of ORCs, \citet{norris21a} searched the 944~MHz radio continuum images from the EMU-PS1 (270 sq deg, rms $\sim$ 27~\ujybm) and found three objects that they labelled an ORC. One (ORC1) surrounded an elliptical galaxy and the other two (ORC2 and ORC3) were associated with a double-lobed AGN. They also found a single ORC (ORC4) surrounding a massive elliptical galaxy in GMRT archival data at 325~MHz. Another single ORC (ORC J0102--2450), the third one around a massive elliptical galaxy, was subsequently discovered in ASKAP data \citep[][]{koribalski21}. A number of candidate ORCs have also since been discovered {  (e.g. \citet{gupta22, lochner23, dolag23, koribalski24}, Koribalski et al., in preparation). } Furthermore, ORC-like radio shell systems around nearby galaxies \citep[e.g., the Physalis system][]{k24-physalis} are adding to the puzzle.

It is now clear {  from MeerKAT observations} that the pair of ORCs (ORC2/3) is associated with the AGN between them { (Macgregor et al., in preparation)}, while the other three  well-studied single ORCs each have a central elliptical host galaxy in the redshift range  $z \sim 0.27-0.55$ \citep{norris21a,koribalski21} implying an ORC diameter of 300 -- 500~kpc. The three single host galaxies are remarkably similar, with stellar masses ($M_{*} \gtrsim 10^{11}\Msun$) and stellar population ages $\gtrsim$1~Gyr \citep{rupke24,coil24}. 

The term ``ORC'' was initially loosely defined, but here we adopt a tighter definition to avoid confusion with other phenomena.  Specifically, we define an ORC as 
an edge-brightened circle
of radio emission, 
without any known corresponding emission at other wavelengths, surrounding a distant galaxy. However, they may also show internal diffuse emission or structure. This definition makes a clear distinction between the single ORCs, which presumably have a common mechanism, and (a) the radio rings which seem to sometimes originate in the relic lobes of a double-lobed radio galaxy { (\citet{norris21a,omar22a}, Macgregor et al., in preparation)}, and (b) the diffuse radio emission seen round some galaxies
\citep[e.g.][]{kumari24,kumari24a}, which may well be related to ORCs but do not show a well-defined ring around a host galaxy.

The mechanism leading to the production of single ORCs is still unclear, but two main classes of model are possible. Most widely favoured is that the rings are the projection of a spherical shell resulting from a shock from the central host galaxy $\sim 10^9$ years ago. This is supported by the brightness profile and by the discovery of a tangential magnetic field in ORC1 \citep{norris22}. The cause of this shock has been variously attributed to a merger of supermassive black holes \citep{norris22}, a merger of galaxies \citep{dolag23},
a  shock wave from a starburst \citep{norris22, coil24}, or the infall of stellar material on to a supermassive black hole \citep{omar22}. Another strong contender is that the rings could result from the collision of a fading ``relic'' radio lobe with an external shock front, re-energising the old electrons \citep{shabala24}.

Detailed studies of the host galaxies might provide clues to distinguish between these models, but optical observations have been frustrated by the faintness of the hosts. A breakthrough was the observation of ORC4 with the Keck telescope by \citet{coil24} who found 3727\AA\ [O\,{\sc ii}] emission with high velocity dispersion extending $\sim$40~kpc from the central galaxy. While this emission is an order of magnitude smaller than the diffuse radio ring emission, it is much larger than expected for a typical early-type galaxy, and may represent a ``smoking gun'' from the event that caused the ORC. 

Here we report the discovery of an ORC, shown in Figure \ref{fig:neworc}, that surrounds a central elliptical galaxy showing significant extended structure, which may represent an even more extreme case of the disturbance noted by \citet{coil24}.
{  Importantly, the new ORC is significantly  smaller and fainter than previously known ORCs}.

\section{MIGHTEE Observations and Data Reduction}

The MIGHTEE survey \citep{jarvis16} is a deep radio survey conducted with the MeerKAT telescope \citep{jonas16}, reaching an rms sensitivity of a few \ujybm as shown in Table  \ref{tab:MeerKAT}.
It covers $\sim$20 \sqdeg\ including the best studied regions of the extragalactic sky: COSMOS \citep{cosmos}, XMM-LSS \citep{xmmlss}, ELAIS-S1 \citep{elais}, and ECDFS \citep{ecdfs}, which includes a deeper region from the LADUMA continuum observations \citep{blyth16}.

The data in this paper are taken from Data Release 1 (DR1) of the MIGHTEE survey \citep{hale24}, which used a total of 564.2 hours of MeerKAT time to obtain deep surveys of the COSMOS, XMM-LSS and CDFS-DEEP fields. 
Further details for each field are given in Table \ref{tab:MeerKAT}.

\begin{table}
{ 
\caption{The search for ORCs in MIGHTEE and EMU continuum images.}
\label{tab:MeerKAT}
\resizebox{\columnwidth}{!}{%
\begin{tabular}{lccccc}
\toprule
Field & Area & median & Beam & Cpts & Cpts \\ 
 && rms & size & &  per deg$^2$ \\
 & [deg$^2$] & [\ujybm] & [\arcsec] & & \\
\midrule
COSMOS & 4.2 & 5.6 & 5.2 & 22420 & 5338 \\
XMM-LSS & 14.4 & 5.1 & 5.0 & 76567 & 5317 \\
CDFS-Deep & 1.5 & 1.9 & 5.5 & 22660 & 15107 \\
\midrule
MIGHTEE Total & 20.1 & & & 121647 & 6052\\
\midrule
EMU-PS1 & 270 & 27 & 15 & 220102 & 815 \\
\bottomrule\\
\end{tabular}%
}
The last two columns give the total number of components (Cpts), and the number of components per square degree, where a component is defined as a Gaussian identified by the survey source finder.
}
\end{table}

Here we give a brief overview of the observations and data processing; further details are given by  \citet{hale24, heywood22}. After flagging the data, the primary and secondary calibrators were used to derive bandpass, delay and time-dependent gain solutions using CASA \citep{casa}. The data were then imaged using WSCLEAN \citep{offringa10}, followed by self-calibration using CUBICAL \citep{kenyon18}, and re-imaging. Strong radio sources were removed using peeling, and the data were then re-imaged with direction-dependent gain errors using DDFACET \citep{tasse23}. The fields were observed over the full bandwidth of 856 -- 1711~MHz, and were processed using two values (0.0 and --1.2) of  Briggs robustness to produce images at two resolutions in each field. In this project, we use the highest resolution images which have a half-power beamwidth of $\sim$5\arcsec, and an effective observing frequency of $\sim$ 1.2 GHz\footnote{  The  effective frequency varies slightly as a function of position, as discussed by \citet{hale24}}.

\subsection{The search for ORCs in MIGHTEE data}

For this project we inspected the survey images by eye, rather than using a catalog of  sources, because source finders can miss the diffuse emission of ORCs and other extended systems. We now describe the process, 
as some techniques are unsuccessful at finding ORCs. The images were loaded into the CARTA visualisation tool \citep{carta}, and scaled so that the  image visible on the screen at any time covered about $10\arcmin \times 10\arcmin$. At this scale, point sources occupy a few pixels of the screen, and a typical ORC ($\sim$1\arcmin\ diameter) is very obvious. The greyscale was typically set to 99\%, so that the noise background is visible, but would sometimes be increased or decreased to examine strong or weak sources. Grid lines were turned on to facilitate scanning the image in a raster scan along lines of constant declination. During the raster scan, the positions of radio ring candidates and related objects were noted. After each field had been scanned in this way, the position of each such object was revisited with a careful examination in radio, optical, and infrared wavelengths.

\section{Results and Discussion}
\subsection{Survey Results}

\begin{figure}
    \includegraphics[width=9cm]{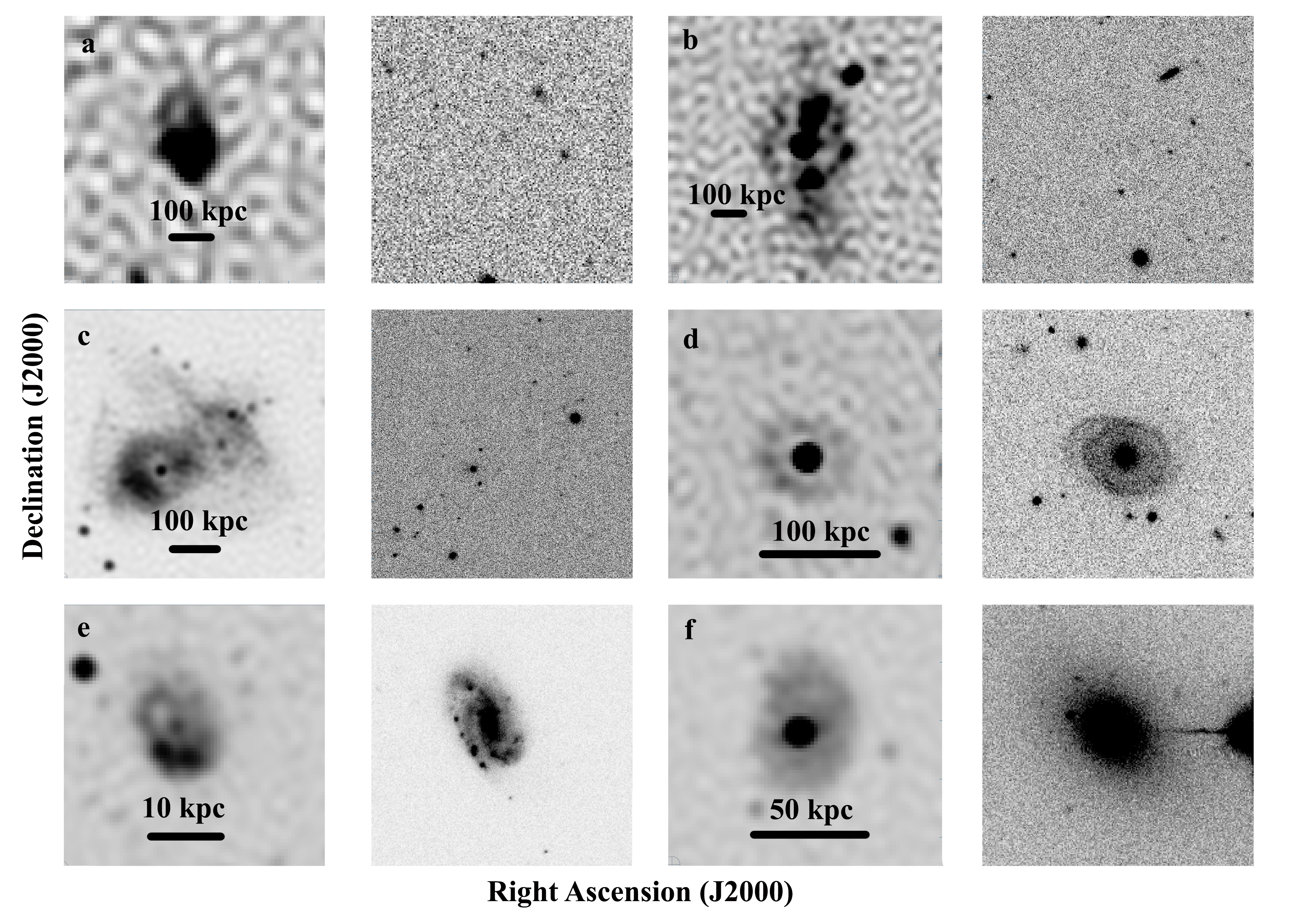}\hfill
    \caption{MIGHTEE 1.2~GHz radio continuum images of sources with rings or ring-like structures,  which might be mistaken for ORCs. Each pair of images show the radio image (left) and the corresponding SDSS $r$-band \citep{sdss} optical image (right). 
    Axis labelling has been omitted in this figure for clarity, but fully labelled images are available in the Supplementary Information.
    \label{fig:candidates} }
\end{figure}

We found 
a wide range of radio loops and ring-like structures in the COSMOS and XMM-LSS fields, but none in the CDFS-DEEP field, {  presumably because of the smaller field size.} Among them are nearby spiral galaxies, S-shaped radio galaxies (RGs), diffuse radio disks and others. We show a few examples of these objects in Table~\ref{tab:candidates} and Figure~\ref{fig:candidates}. {  For example, (b) and (c) show some circular structure, but this is clearly part of a larger complex extended emission region, including in (c) a ring that does not surround a galaxy. We refer to such objects as ``ORC candidates''.} Only one object fitted the definition of an ORC as discussed above, and we now focus on the properties of this object: ORC~J0219--0505.

\begin{table}
{ 
\caption{Radio sources with ring-like structures found in the MIGHTEE 1.2~GHz radio continuum images.
Their radio and SDSS $r$-band images are shown in Fig.~\ref{fig:candidates}.  }
\label{tab:candidates}
\resizebox{\columnwidth}{!}{
\begin{tabular}{llll}
\toprule
Label & RA, Dec(J2000) 
& Redshift 
& Description\\ 
& (degrees) & \\
\midrule
a      &   33.5167, –3.8661     &  
0.9770&     radio loop associated with RG/QSO          \\
b      &   33.6521, –5.0858     &  
0.6360&     galaxy in a cluster                                \\
c      &   34.9237, –4.0097     &  
0.2042&     radio ring around E-type galaxy \\
&&& plus diffuse emission   \\
d      &   35.2792, –4.9991     &  
0.1395&     face-on spiral galaxy  with SF ring                  \\
e      &   35.8050, –4.5194     &  
0.0176&     nearby spiral galaxy                                 \\
f      &   36.7458, –4.5903     &  
0.0697&     elliptical galaxy with diffuse radio disk \\

\bottomrule\\
\end{tabular}%
}
}
References for redshifts: (a) \citet{ozdes}, (b) Legacy Survey, (c,d) \citet{baldry18}, (e) \citet{ahumada20}, (f) \citet {cheng21}
\end{table}

\begin{table}
{ 
\caption{Radio properties of the MIGHTEE ORC and its host galaxy}
\label{tab:radio}
\resizebox{\columnwidth}{!}{%
\begin{tabular}{lcccc}
\toprule
\multicolumn{1}{c}{Property} & ORC & Host  & Ref. \\ 
\midrule
RA, degrees (J2000) & 34.80058    & 34.80175\\
Dec, degrees (J2000) & --5.08406 &  --5.08406  \\
1.2~GHz integrated flux density ($\mu$Jy)  &  $1079 \pm 108$ & $171 \pm 18$ &  (1)\\
1.2~GHz peak flux density ($\mu$Jy/bm) & $48 \pm 7$ & $146 \pm 15$ &  (1)\\
1.4 GHz flux density & -- & $223 \pm 37$ & (2) \\
325 MHz flux density ($\mu$Jy) & -- &{$1512 \pm 332$} & (3) \\
144 MHz flux density ($\mu$Jy) & -- &{$2000 \pm 500$} & (4) \\
spectral index (4-point least-squares fit) & & {$-1.22 \pm 0.11$} & \\
\bottomrule\\
\end{tabular}%
}
}

The MeerKAT peak and integrated flux densities of the ORC,were measured using CARTA \citep{carta}. The integrated flux density of the ORC was measured by manually drawing a region around the source and using the CARTA statistics tool, and excludes the compact radio sources associated with the host and C1. No background has been subtracted.  The position of the ORC is the centre of the radio ring. --- References: (1) MeerKAT this paper, (2) VLA \citet{simpson06}, (3) GMRT \citet{singh14}. (4) LOFAR \citet{hale19} 
\end{table}


\subsection {Radio properties of ORC J0219--0505}
\label{discovery}

\begin{figure*}
\includegraphics[width=18cm]{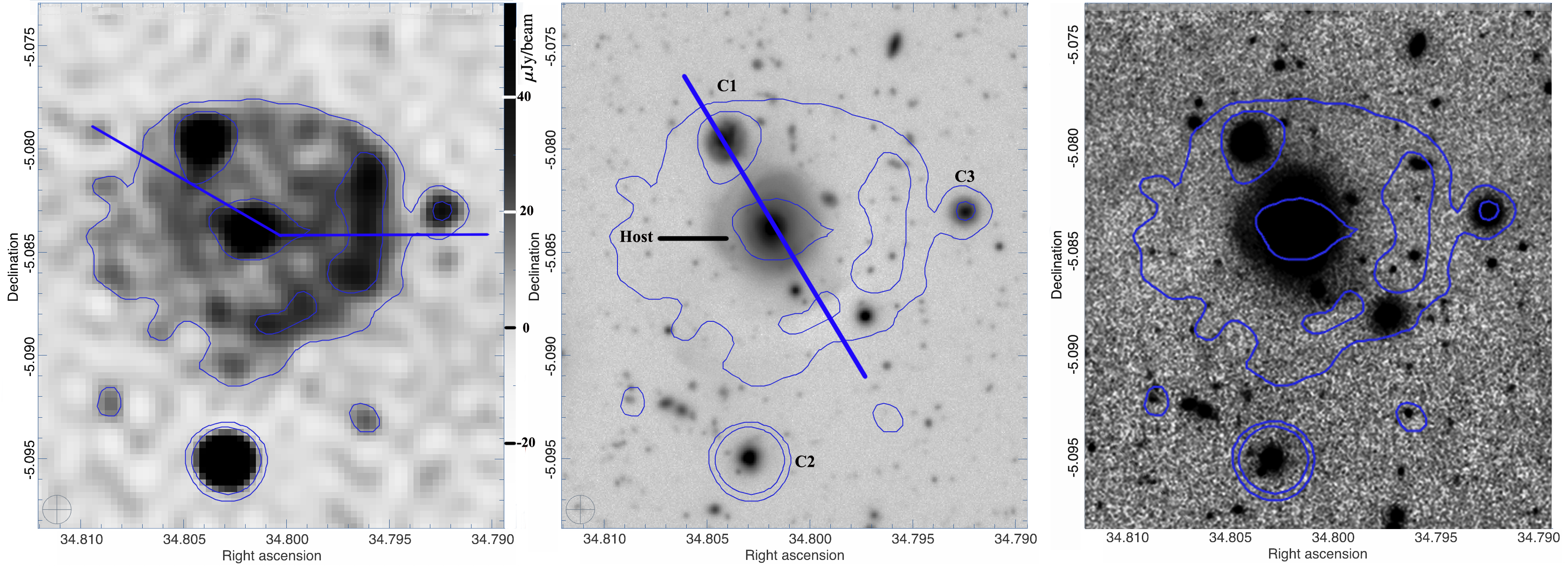}\hfill
    \caption{{  Left:} MIGHTEE 1.2~GHz radio continuum image of ORC J0219--0505 (resolution 5\arcsec). The central radio source is associated with the host elliptical galaxy at $z = 0.196$. Diffuse radio emission surrounds the host forming a 35\arcsec\ diameter ring. Two compact radio sources, marked as C1 and C2 {  in the centre panel}, are associated with companion galaxies, while C3 is an unassociated background galaxy. Contours at 8 and 24 $\mu$Jy/beam are chosen to show the overall extent and the position of the peaks in the ring. The colourbar is in units of $\mu$Jy/beam. {  Centre:} Subaru HSC $R$-band image \citep{subaru}, showing extended optical emission around the host galaxy, including a shell to the east, and a sharp edge to the southwest. --- {  Right:} VIDEO $K_{\rm s}$-band infrared image \citep{jarvis2013} with a transfer function chosen to emphasise the faint emission, showing the same extended emission and the sharp edge. The amplitude profile along the line on the centre panel is shown in Figure \ref{fig:rband_profile}.
    \label{fig:3bands}}
\end{figure*}

The MIGHTEE image of the new ORC (ORC J0219--0505) is shown in Figures \ref{fig:neworc} and \ref{fig:3bands}. It consists of an edge-brightened ring of radio emission surrounding a compact radio source in the centre. The ring is filled with faint diffuse emission, with hints of some structure, and there is also a marginal detection of some faint diffuse emission extending to the south-east of the ring.
The radio ring has a diameter of 35\arcsec, corresponding to 114~kpc at a redshift of $z_{\rm spec} = 0.196$.

\begin{figure}
\includegraphics[width=8cm]{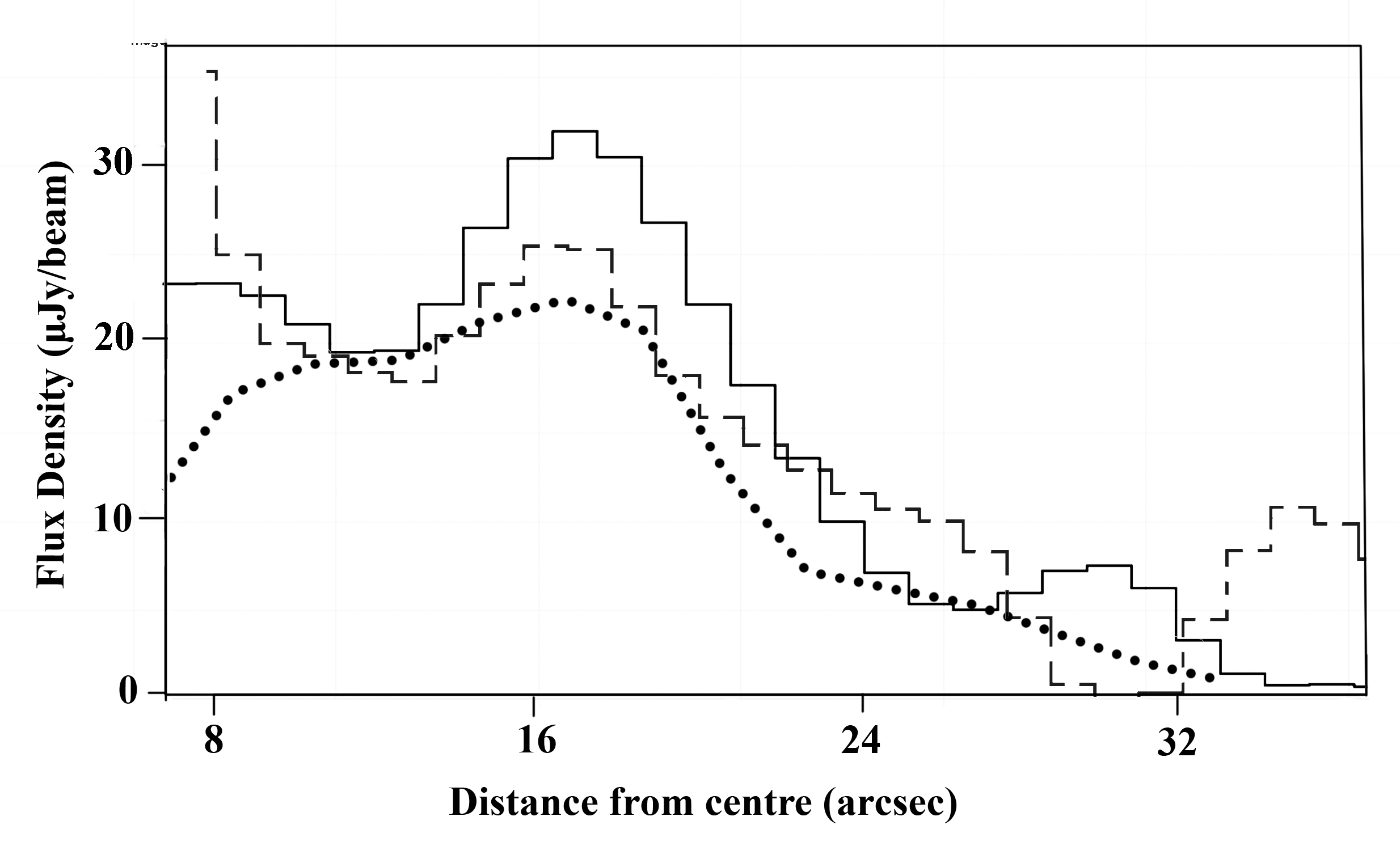}\hfill
    \caption{The radial profile of the radio emission in ORC J0219--0505, showing the edge-brightened ring surrounded by a sharp drop in flux density. {  The solid line shows the flux density along a line extending west from the centre of the ring, and the dashed line shows the flux density along a line extending at position angle 60\degr from the centre, chosen to avoid continuum sources, and  shown as blue lines in the left hand panel of Figure \ref{fig:3bands}. The dotted line show the normalised integrated radio emission in a 2\arcsec\ wide annulus, divided by the annulus area in arcsec$^2$, with the flux density of the host and companion galaxies subtracted. No background has been subtracted. 
    The annuli, and the origin of the lines, are centred on a position estimated to be the centre of the ORC ring, as shown in Table \ref{tab:radio}, which is about 4\arcsec\ to the west of the radio core. For the dotted line, the x axis shows  the mean radius of each annulus. }
    The flux density uncertainty is estimated to be 10\%.}
    \label{fig:radio_profile}
\end{figure}

{ 
In Figure \ref{fig:radio_profile} we plot the radial profile of the ring. The one-dimensional profile shows  a maximum at the position of the ring that has an approximately Gaussian profile with a full width at half maximum of $\sim$9\arcsec, or $\sim$30 kpc, so is clearly resolved by the 5\arcsec\ beamwidth. This is consistent with both the \citet{dolag23} and \citet{shabala24} models, but may play a role in future modelling.

We note that the centre of the radio ring appears offset by about 4\arcsec\
to the west of the radio continuum source and the host galaxy. This is naturally explained as an orientation effect in the \citet{shabala24} model, but in the \citet{dolag23} model would require an off-centre origin of the shock that caused the shell, such as a galaxy-galaxy interaction. It would be difficult to explain this offset in a model based on the central super-massive black hole (SMBH).
}

The new ORC is about half the angular size and one fifth of the typical integrated flux density of the previously found ORCs. When convolved to the 15\arcsec\ resolution of  EMU-PS1, the ring appears as a fuzzy disc with marginal evidence for a ring, making it likely that a similar ring would have been missed when searching the ASKAP or GMRT images. The ORC properties are given in Table~3.

\begin{figure}
\includegraphics[width=8cm]{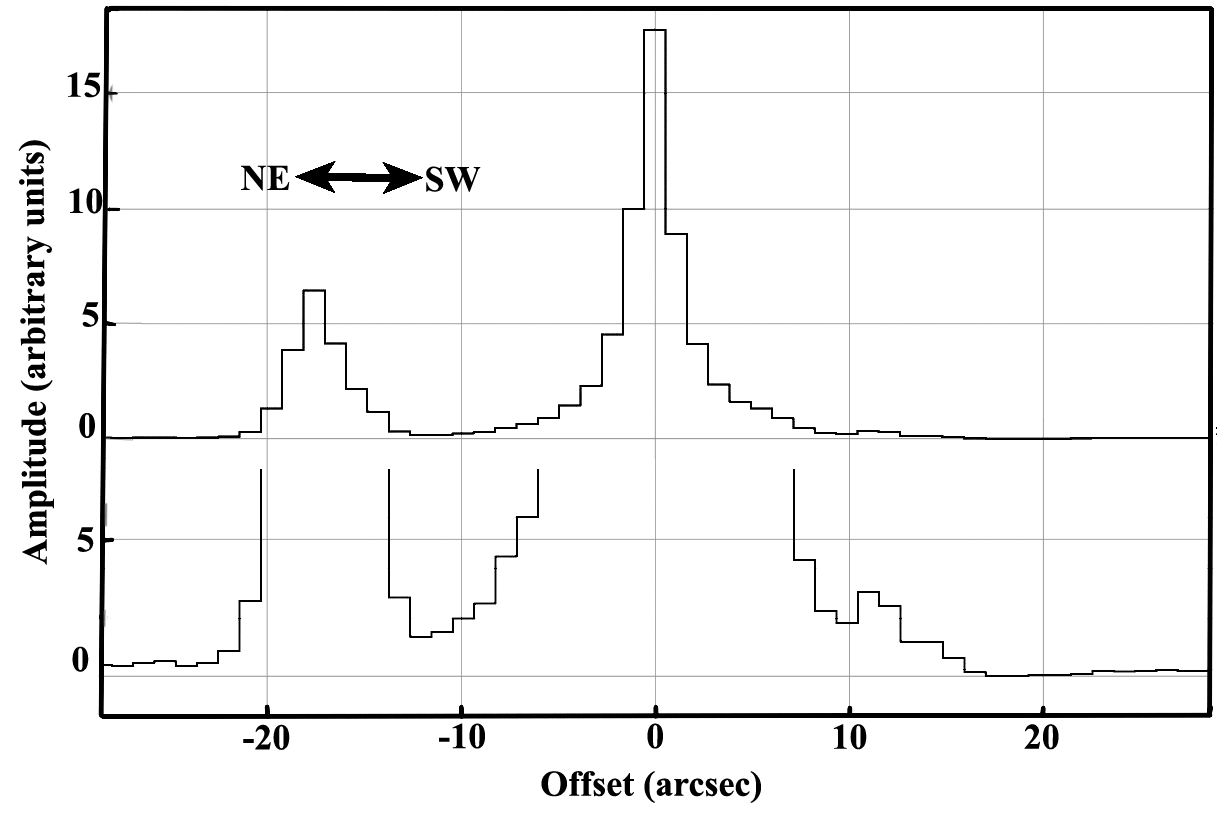}\hfill
    \caption{The Subaru HSC $R$-band amplitude profile across ORC J0219--0505 along the line on the centre panel of Figure \ref{fig:3bands}. The position angle of the line was chosen to show the extended emission up to 15\arcsec\ to the south-west, and also the extended emission between the host and C1.}
    \label{fig:rband_profile}
\end{figure}

\subsection {Multiwavelength properties of ORC J0219--0505}
The central radio source of the ORC is coincident with the bright elliptical galaxy 
WISEA J021912.43--050501.8 which has a spectroscopic redshift of $z_{\rm spec} = 0.196$ \citep{simpson06,baldry18}.  Its 1.2~GHz radio emission (0.15 mJy, or 1.6 $\times 10^{22}$ W/Hz, see Table~3) is probably generated by an AGN. \citet{omori23} derive a stellar mass of $M_{*} = 10^{11.73}\Msun$ and an effective radius of $R_{\rm eff}$ = 3\arcsec, and suggest, based on a supervised deep learning model, that this is probably a merger.

Within the north-east quadrant of the ring is a compact source (labelled as C1 in Figure \ref{fig:3bands}) which is spatially coincident with the spiral galaxy 
WISEA J021912.96--050446.5 at $z_{\rm spec} = 0.201$ \citep{baldry18}. 
This galaxy appears blue in optical images, and has WISE \citep{wise} colours (W1--W2=0.54, W2--W3=3.79) which suggest it is a starburst, LIRG, or Seyfert galaxy. Its 1.2 GHz radio emission (0.15 mJy, or 1.6 $\times 10^{22}$ W\,Hz$^{-1}$) corresponds to a plausible star formation rate of $\sim$ 10\Msun\,yr$^{-1}$, so it is unclear whether this source has an AGN. Its redshift places it at a relative velocity\footnote{using the equation $v = c\delta z/(1+z)$} of $\sim$1300~km\,s$^{-1}$ with respect to the host galaxy. This velocity is high for a pair of interacting galaxies, unless they are in a rich cluster, but implausibly low for a chance association \citep[e.g.][]{ferragamo21}. It is thus likely this galaxy is  interacting with the  ORC host. 

There is extensive deep optical and IR data available for this field, two examples of which are shown in Figure \ref{fig:3bands}. Both the host and C1 companion are also detected  at 250 $\mu$m by Herschel-SPIRE \citep{spire}. 
However, like other ORCs, the ring has no corresponding diffuse emission in any of the  available optical or IR observations. No clear counterpart of either the ring, the host, or the companion, is seen in XMM data, and the region lies outside available eROSITA and Chandra data. However, it is part of the cluster SXDF88XGG and within the contours of the extended diffuse X-ray emission associated with that cluster \citep{finoguenov10}.

A second, radio-bright companion galaxy 
WISEA J021912.70--050541.9 \citep[$z_{\rm spec} = 0.194$,][]{simpson06,baldry18} (labelled as C2 in Figure \ref{fig:3bands}) is found 40\arcsec\ south of the host, but there is no indication of any interaction between this galaxy and the host.

A third radio source (labelled as C3 in Figure \ref{fig:3bands}) is a background source (GAMA J021910.18--050458.8 at $z = 0.431$ \citep{baldry18}) that has no physical connection to this system.

\subsection {Extended optical emission in the ORC host galaxy}
Figures \ref{fig:3bands} and \ref{fig:rband_profile} show extended optical emission features extending as a far as 15\arcsec, or 50~kpc, from the centre of the galaxy, compared to the 3\arcsec\ (10~kpc) half-width of the galaxy. These extended features include a faint bridge  (seen in Figures \ref{fig:3bands} and \ref{fig:rband_profile} as non-zero emission between the host and C1) between the host galaxy and the companion galaxy C1. These features are seen in all available optical/NIR bands (only two of which are shown here for reasons of space) but are brightest in $R$-band, and so are presumably due to starlight rather than, for example, dust or emission lines. Such extended features are characteristic of disturbance by a  merger or interaction, and we speculate that the companion galaxies may have participated in this interaction.
This supports ORC models in which the ORC is a shell of radio emission resulting from energised electrons energised by a spherical shock from a  galaxy merger \citep{dolag23,koribalski24} or SMBH merger. 

Figure \ref{fig:3bands} also shows a sharp edge to this extended emission to the south west of the galaxy, coincident with the inner edge of the shell of radio emission. 
This edge is seen in both the Subaru $R$-band \citep{subaru} and the VISTA Deep Extragalactic Observations (VIDEO) $K_{\rm s}$-band \citep{jarvis2013} images, making it unlikely to be an artefact. This suggests the diffuse extended starlight may  be affected by  extinction by dust associated with the shock. Detailed modelling of these observations will be discussed in a future paper. The extended structures are also reminiscent of those found by \citet{coil24} in the host galaxy of ORC~4, and spectroscopic observations will be key to understanding this source.

\subsection{The flux density distribution of ORCs}

The ORCs found so far \citep{norris21a, koribalski21} are remarkably uniform, with integrated 1--GHz flux densities of 4 -- 10 mJy, angular sizes of $\sim$ 1 arcmin, redshifts of 0.25 -- 0.55, and linear sizes of 300 -- 500 kpc. 
This raises the question of whether the ORC phenomenon is limited to a particular range of parameters by physical constraints, or whether the apparent uniformity is due to a spatial filter effect imposed by the observations. For example, why do we not find any larger or lower redshift ORCs?

Table \ref{tab:MeerKAT} shows that the MIGHTEE survey has about one fifteenth the area of EMU-PS1, but is about ten times more sensitive, so that MIGHTEE finds many more sources per square degree than EMU-PS1. If ORCs occupy  the narrow range of parameters listed above, then, ignoring resolution effects, simply scaling the number found in EMU-PS1 \citep{norris21b} by survey area would predict that we would find $\sim$ 0.1 ORCs in the MIGHTEE observations. If, on the other hand, the flux density distribution of ORCs mirrors that of the extragalactic source population, then we would expect the ratio of the number of ORCs to the number of other extragalactic sources to be roughly constant. Thus, scaling by the number of Gaussians suggests we might expect $\sim$0.5 ORC in the MIGHTEE data, {  which is consistent with the one ORC we have found in MIGHTEE. Furthermore, this new ORC is significantly fainter and smaller than previously known ORCs, suggesting that the  apparent uniformity of ORC properties was primarily due to selection effects. 

Importantly,}
the discovery of one ORC in MIGHTEE is  consistent with the {  flux density distribution} of ORCs mirroring the  flux density distribution of the extragalactic source population. This also suggests that there is a fainter population of ORCs that  remains to be discovered in MeerKAT and other data.

{ 
\subsection{Implications for ORC models}
While the discovery of an ORC in the MeerKAT data cannot itself distinguish between the competing models for ORCs, it does introduce some important constraints.

First, this new ORC is significantly smaller and fainter than previously known ORCs, suggesting that the apparent uniformity of their properties was a selection effect. In particular the new result is consistent with ORCs having a similar flux density distribution similar to that of the broader extragalactic radio source population.

Second, whilst the size of the 520 kpc shell diameter of  ORC1 is consistent with the launch of a spherical shock from an merger or starburst event $\sim$ 1 Gyr ago \citep{norris22}, this becomes more difficult to model for the 114 kpc diameter of the MeerKAT ORC shell.

Third, the extended optical/IR emission of the host galaxy indicates a disturbance in the past, thus naturally supporting the \citet{dolag23} galaxy merger model. However, this disturbance might also generate the shock required by the \citet{shabala24} model.

Finally, the offset between the host galaxy and the centre of the ring is naturally accounted for by the \citet{shabala24} model, can be accommodated by an asymmetric merger in the \citet{dolag23} model, but is difficult to accommodate in any model focusing on the SMBH.

}

\section{Conclusion and Further Work}

We have discovered one new ORC (ORC J0219--0505) in the deep MIGHTEE survey of the XMM-LSS field, increasing the number of well-defined single ORCs from three to four. ORC J0219--0505 has a diameter of 35\arcsec, corresponding to 114~kpc at the redshift ($z_{\rm spec} = 0.196$) of the elliptical host galaxy. We measure a ring width of $\sim$10\arcsec, ie. $\sim$33~kpc. The new ORC is a factor 2--3 smaller and somewhat fainter than its predecessors discovered with ASKAP and GMRT, {  implying that the uniform properties of previously known ORCs are primarily due to selection effects, such as sensitivity and resolution}. Its distinct radio ring and central radio source could not have been seen in those earlier surveys, primarily because it would be below their resolution limits.

The host galaxy of the new ORC is a massive ($M_{*} = 10^{11.73}$\Msun) elliptical galaxy with extended 
features visible in optical and infrared images. These features extend to $\sim$ 50 kpc, almost to the radio ring of the ORC. There is also a companion spiral galaxy ($z_{\rm spec} = 0.201$) located in the ring of the ORC, and a BCD galaxy ($z_{\rm spec} = 0.194$) to the south, and  both may be interacting with the MIGHTEE ORC host galaxy. This extended structure supports models of ORCs in which the ORC phenomenon is triggered by a galaxy merger \citep{dolag23,koribalski24} or a SMBH merger. {  On the other hand, the offset between the host galaxy and the centre of the ring is most easily explained by the \citet{shabala24} model.}

{  The discovery of the MIGHTEE ORC, which is much smaller and fainter than the previously discovered single ORCs, suggests their similarity is primarily due to selection effects. Its discovery in the 20 \sqdeg\ MIGHTEE field is also consistent with ORCs having a  flux density distribution similar to that of   the general population of extragalactic radio sources. }

It is of prime importance to understand  the extended morphological features visible at optical wavelengths, as they may hold a key to understanding the process that generates ORCs. This will be the subject of further work, {  which will also explore the polarisation and spectral properties of this ORC.}

\section* {Data Availability}
The MIGHTEE DR1 data on which this paper is based are available from \url{https://doi.org/10.48479/7msw-r692} \citep{hale24}.

\section*{Acknowledgements}

We thank the referee, Stas Shabala, for valuable comments which have significantly enhanced this paper.

MJJ acknowledge the support of a UKRI Frontiers Research Grant [EP/X026639/1], which was selected by the European Research Council, and the STFC consolidated grants [ST/S000488/1] and [ST/W000903/1].
MJJ and CLH also acknowledge support from the Oxford Hintze Centre for Astrophysical Surveys which is funded through generous support from the Hintze Family Charitable Foundation.

\subsection*{Meerkat}
The MeerKAT telescope is operated by the South African Radio Astronomy Observatory, which is a facility of the National Research Foundation, an agency of the Department of Science and Innovation. 

\subsection*{Ilifu}
We acknowledge the use of the ilifu cloud computing facility – www.ilifu.ac.za, a partnership between the University of Cape Town, the University of the Western Cape, Stellenbosch University, Sol Plaatje University and the Cape Peninsula University of Technology. The Ilifu facility is supported by contributions from the Inter-University Institute for Data Intensive Astronomy (IDIA – a partnership between the University of Cape Town, the University of Pretoria and the University of the Western Cape, the Computational Biology division at UCT and the Data Intensive Research Initiative of South Africa (DIRISA). The authors acknowledge the Centre for High Performance Computing (CHPC), South Africa, for providing computational resources to this research project.

\subsection*{Hyper Suprime-Cam}
The Hyper Suprime-Cam
(HSC) collaboration includes the astronomical communities of Japan and Taiwan, and Princeton University. The HSC instrumentation and software were developed by the National Astronomical Observatory of Japan (NAOJ), the Kavli Institute for the Physics and Mathematics of the Universe (Kavli IPMU), the University of Tokyo, the High Energy Accelerator Research Organization (KEK), the Academia Sinica Institute for Astronomy and Astrophysics in Taiwan (ASIAA), and Princeton University. Funding was contributed by the FIRST program from Japanese Cabinet Office, the Ministry of Education, Culture, Sports, Science and Technology (MEXT), the Japan Society for the Promotion of Science (JSPS), Japan Science and Technology Agency (JST), the Toray Science Foundation, NAOJ, Kavli IPMU, KEK, ASIAA, and Princeton University.

\subsection*{VIDEO/UltraVISTA}
This work is based on data products from observations made with ESO Telescopes at the La Silla Paranal Observatory under ESO programme ID 179.A-2005 (Ultra-VISTA) and ID 179.A-2006(VIDEO) and on data products produced by CALET and the Cambridge Astronomy Survey Unit on behalf of the Ultra-VISTA and VIDEO consortia.

\subsection*{CFHTLS}
Based on observations obtained with MegaPrime/MegaCam, a joint project of CFHT and CEA/IRFU, at the Canada-France-Hawaii Telescope (CFHT) which is operated by the National Research Council (NRC) of Canada, the Institut National des Science de l’Univers of the Centre National de la Recherche Scientifique (CNRS) of France, and the University of Hawaii. This work is based in part on data products produced at Terapix available at the Canadian Astronomy Data Centre as part of the Canada-France-Hawaii Telescope Legacy Survey, a collaborative project of NRC and CNRS. 

\subsection*{SDSS}
Funding for the Sloan Digital Sky Survey V has been provided by the Alfred P. Sloan Foundation, the Heising-Simons Foundation, the National Science Foundation, and the Participating Institutions. SDSS acknowledges support and resources from the Center for High-Performance Computing at the University of Utah. SDSS telescopes are located at Apache Point Observatory, funded by the Astrophysical Research Consortium and operated by New Mexico State University, and at Las Campanas Observatory, operated by the Carnegie Institution for Science. The SDSS web site is \url{www.sdss.org}.

SDSS is managed by the Astrophysical Research Consortium for the Participating Institutions of the SDSS Collaboration, including Caltech, The Carnegie Institution for Science, Chilean National Time Allocation Committee (CNTAC) ratified researchers, The Flatiron Institute, the Gotham Participation Group, Harvard University, Heidelberg University, The Johns Hopkins University, L’Ecole polytechnique f\'{e}d\'{e}rale de Lausanne (EPFL), Leibniz-Institut f\"{u}r Astrophysik Potsdam (AIP), Max-Planck-Institut f\"{u}r Astronomie (MPIA Heidelberg), Max-Planck-Institut f\"{u}r Extraterrestrische Physik (MPE), Nanjing University, National Astronomical Observatories of China (NAOC), New Mexico State University, The Ohio State University, Pennsylvania State University, Smithsonian Astrophysical Observatory, Space Telescope Science Institute (STScI), the Stellar Astrophysics Participation Group, Universidad Nacional Aut\'{o}noma de M\'{e}xico, University of Arizona, University of Colorado Boulder, University of Illinois at Urbana-Champaign, University of Toronto, University of Utah, University of Virginia, Yale University, and Yunnan University.

\subsection*{Legacy Surveys Sky Viewer}
The Legacy Surveys consist of three individual and complementary projects: the Dark Energy Camera Legacy Survey (DECaLS; Proposal ID \#2014B-0404; PIs: David Schlegel and Arjun Dey), the Beijing-Arizona Sky Survey (BASS; NOAO Prop. ID \#2015A-0801; PIs: Zhou Xu and Xiaohui Fan), and the Mayall $z$-band Legacy Survey (MzLS; Prop. ID \#2016A-0453; PI: Arjun Dey). DECaLS, BASS and MzLS together include data obtained, respectively, at the Blanco telescope, Cerro Tololo Inter-American Observatory, NSF’s NOIRLab; the Bok telescope, Steward Observatory, University of Arizona; and the Mayall telescope, Kitt Peak National Observatory, NOIRLab. Pipeline processing and analyses of the data were supported by NOIRLab and the Lawrence Berkeley National Laboratory (LBNL). The Legacy Surveys project is honored to be permitted to conduct astronomical research on Iolkam Du’ag (Kitt Peak), a mountain with particular significance to the Tohono O’odham Nation.

NOIRLab is operated by the Association of Universities for Research in Astronomy (AURA) under a cooperative agreement with the National Science Foundation. LBNL is managed by the Regents of the University of California under contract to the U.S. Department of Energy.

This project used data obtained with the Dark Energy Camera (DECam), which was constructed by the Dark Energy Survey (DES) collaboration. Funding for the DES Projects has been provided by the U.S. Department of Energy, the U.S. National Science Foundation, the Ministry of Science and Education of Spain, the Science and Technology Facilities Council of the United Kingdom, the Higher Education Funding Council for England, the National Center for Supercomputing Applications at the University of Illinois at Urbana-Champaign, the Kavli Institute of Cosmological Physics at the University of Chicago, Center for Cosmology and Astro-Particle Physics at the Ohio State University, the Mitchell Institute for Fundamental Physics and Astronomy at Texas A\&M University, Financiadora de Estudos e Projetos, Fundacao Carlos Chagas Filho de Amparo, Financiadora de Estudos e Projetos, Fundacao Carlos Chagas Filho de Amparo a Pesquisa do Estado do Rio de Janeiro, Conselho Nacional de Desenvolvimento Cientifico e Tecnologico and the Ministerio da Ciencia, Tecnologia e Inovacao, the Deutsche Forschungsgemeinschaft and the Collaborating Institutions in the Dark Energy Survey. The Collaborating Institutions are Argonne National Laboratory, the University of California at Santa Cruz, the University of Cambridge, Centro de Investigaciones Energeticas, Medioambientales y Tecnologicas-Madrid, the University of Chicago, University College London, the DES-Brazil Consortium, the University of Edinburgh, the Eidgenossische Technische Hochschule (ETH) Zurich, Fermi National Accelerator Laboratory, the University of Illinois at Urbana-Champaign, the Institut de Ciencies de l’Espai (IEEC/CSIC), the Institut de Fisica d’Altes Energies, Lawrence Berkeley National Laboratory, the Ludwig Maximilians Universitat Munchen and the associated Excellence Cluster Universe, the University of Michigan, NSF’s NOIRLab, the University of Nottingham, the Ohio State University, the University of Pennsylvania, the University of Portsmouth, SLAC National Accelerator Laboratory, Stanford University, the University of Sussex, and Texas A\&M University.

BASS is a key project of the Telescope Access Program (TAP), which has been funded by the National Astronomical Observatories of China, the Chinese Academy of Sciences (the Strategic Priority Research Program “The Emergence of Cosmological Structures” Grant \# XDB09000000), and the Special Fund for Astronomy from the Ministry of Finance. The BASS is also supported by the External Cooperation Program of Chinese Academy of Sciences (Grant \# 114A11KYSB20160057), and Chinese National Natural Science Foundation (Grant \# 12120101003, \# 11433005).

The Legacy Survey team makes use of data products from the Near-Earth Object Wide-field Infrared Survey Explorer (NEOWISE), which is a project of the Jet Propulsion Laboratory/California Institute of Technology. NEOWISE is funded by the National Aeronautics and Space Administration.

The Legacy Surveys imaging of the DESI footprint is supported by the Director, Office of Science, Office of High Energy Physics of the U.S. Department of Energy under Contract No. DE-AC02-05CH1123, by the National Energy Research Scientific Computing Center, a DOE Office of Science User Facility under the same contract; and by the U.S. National Science Foundation, Division of Astronomical Sciences under Contract No. AST-0950945 to NOAO.

\subsection*{Legacy Survey Photometric Redshifts}
The Photometric Redshifts for the Legacy Surveys (PRLS) catalog used in this paper was produced thanks to funding from the U.S. Department of Energy Office of Science, Office of High Energy Physics via grant DE-SC0007914.

\bibliographystyle{mnras}
\bibliography{main} 

\bsp	
\label{lastpage}
\end{document}